# Review of Poher experiment on fields produced by electric discharges in a superconductor

R.A. Lewis[*]
The Pennsylvania State University, University Park, Pennsylvania 16802

**Abstract**
High current through cold YBaCuO produced effects on the mounting and on remote piezoelectric detectors. The effects were not observed with warm YBaCuO or other cold or warm materials. Two hypotheses are evaluated for the direct mechanical forces: a hypothetical vacuum field, and a model in which boil off of liquid nitrogen acts as a thruster. Data from the remote detector are compared with data from Podkletnov. Properties of a field required to explain data with the remote detectors are described.

## Overview of Poher experiments

Experiments[1] described in this article were performed in 2006 in a private laboratory in Toulouse, France. A capacitor charged to several kilovolts was discharged through various targets, including a two-layer YBaCuO disc.

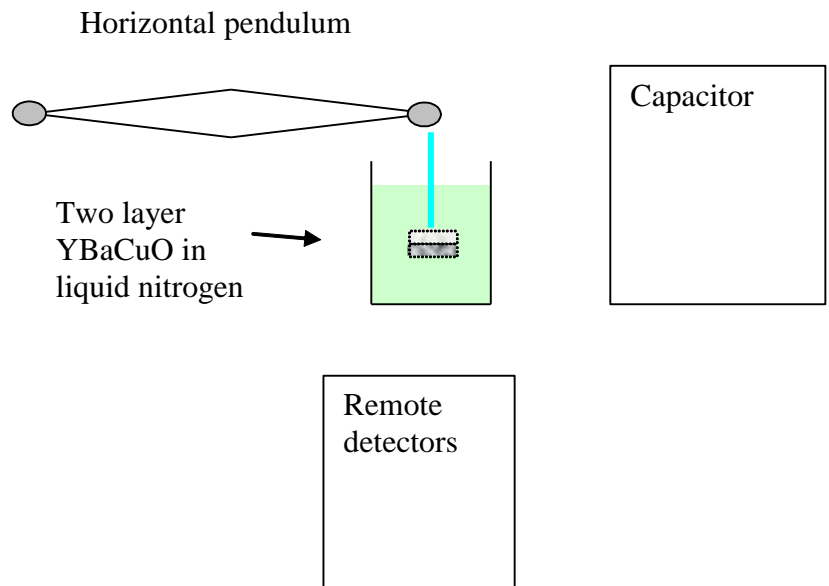

Fig. 1. Layout of capacitor, target and detectors.
Momentum produced by forces on the target mount was transmitted to a horizontal pendulum. Disturbances extending beyond the confines of the target mount were measured with remote detectors located beneath the target.

## Comments on Universon theory

The explanation favored by Poher is a vacuum field termed[2] the Universon. For brevity in this review the field is denoted with $\Upsilon$, the Greek upsilon. The theory is described in a supplement[3] to the Poher article.

---

[*] Retired.

Particles continually absorb and emit Universon particles. For example an electron,
$$\Upsilon + e^- \leftrightarrow e^{*-} \qquad \text{eqn. 1.}$$
The $\Upsilon$ particle in equation 1 has an energy of $8.0*10^{-21}$ Joules, and travels at the speed of light. The lifetime of the excited state $e^{*-}$ is $7.8*10^{-14}$ seconds. If the particle is moving at constant velocity, the $\Upsilon$ absorption and emission result in fluctuations in the mass of the particle. However if the particle is accelerating, the $\Upsilon$ particles emitted have different momentum than the $\Upsilon$ absorbed. The net result is a flux of $\Upsilon$ particles from an accelerating source.

Poher states that the two layer YBaCuO superconductor used results in a long free path for electrons to accelerate. Hence the bi-layer superconductor produces a measurable flux of $\Upsilon$ particles, which are detected by a piezoelectric accelerometer.

The mechanical properties of the $\Upsilon$ beam are treated as an energy/momentum relation E=pc, appropriate for a beam moving at the speed of light. The possibility of a different relation, such as the near field of an electromagnetic field, is beyond the development of the theory. Poher hypothesizes that the Universon gains energy from the vacuum.

The apparatus is termed a propelling device, in a patent application[4].

## Thermodynamics for liquid nitrogen hypothesis

Poher briefly mentions the possibility that heating of liquid nitrogen might contribute to the motion of the ceramic support. In this section the nitrogen heating is treated as an ideal thruster, to evaluate how much the mechanism could possibly contribute to motion of the ceramic support.

Suppose that N molecules of nitrogen are heated from $T_0=77$ k to some temperature T. The energy required to heat the nitrogen is given by
$$E_{N2} = \tfrac{3}{2} NkT + N \cdot E_{latent} - \tfrac{3}{2} NkT_0 \qquad \text{eqn. 2,}$$
where $E_{latent} = 199 \, J/g \cdot 4.65 \bullet 10^{-23} \, g/N_2$ is the latent heat of evaporation per $N_2$ molecule. Assuming that the gas escapes in one direction, the momentum carried by the gas is
$$p_{N2} = Nmv = N\sqrt{3mkT} \qquad \text{eqn. 3,}$$
where $m=4.65*10^{-26}$ kg is the mass of an $N_2$ molecule, and $k=1.38*10^{-23}$ J/k is the Boltzmann constant.

The minimum energy to momentum ratio is achieved by heating the nitrogen to a temperature of T = $2E_{latent}/3k - T_0 = 372$ kelvin. At this temperature, the mean thermal velocity is 575 m/sec. The energy versus temperature curve has a shallow minimum, varying by 25% over the range $77 < T < 1900$ k.

Assuming the minimum energy to momentum ratio, the energy required to vaporize and heat the LN2 to 372 k is 330 Joules/gram. The energy to momentum ratio is 574 m/sec, comparable to the mean thermal velocity.

## Local mechanical forces from target

Forces produced in a target are transmitted mechanically to a horizontal pendulum. The motion of the pendulum is a measure of the momentum produced.

The diamond shape horizontal pendulum had a total mass of 794±1 g and a length of 588±0.5 mm. It was made of an aluminum alloy and had lead masses (M=320 g) at its two tips. The right tip received a supplementary mass (m=1.498±0.002 g) after careful adjustment. The masses on the horizontal pendulum were chosen to match the mass of the ceramic support, for optimum energy and momentum transfer.

Ignoring the mass distributed along the length of the pendulum, the energy and momentum transferred to the pendulum are related to the height h reached by the tip, and the initial velocity V at the tip,

$$E = \tfrac{1}{2}MV^2 + \tfrac{1}{2}MV^2 + \tfrac{1}{2}mV^2 = Mgh - Mgh + mgh \qquad \text{eqn. 4.}$$

$$p = MV + MV + mV = 0.137 \frac{kg \cdot m}{s}\left(\frac{h}{1m}\right)^{1/2} \qquad \text{eqn. 5.}$$

Poher expresses the momentum in mixed units, and takes into account the distributed mass,

$$p = 7.3 \frac{g \cdot m}{s}\left(\frac{h}{1mm}\right)^{1/2} \qquad \text{eqn. 6.}$$

Data are presented on how far the pendulum moves under various conditions, such as varying voltage and targets. Effects were largest using a cold, two layer YBaCuO target. No motion of the pendulum was detected using warm YBaCuO, cold or warm aluminum. Data taken with the cold YBaCuO target are presented graphically. The propulsive momentum appears to vary as the square of the discharge voltage. A typical data point is p=36 g m/s at U=2000 volts, corresponding to h=24.3 mm using eqn. 6. To analyze the data in terms of energy transfer, approximate the relation between height h and capacitor voltage as follows:

$$h = 24.3mm\left(\frac{U}{2000 volts}\right)^4 \qquad \text{eqn. 7.}$$

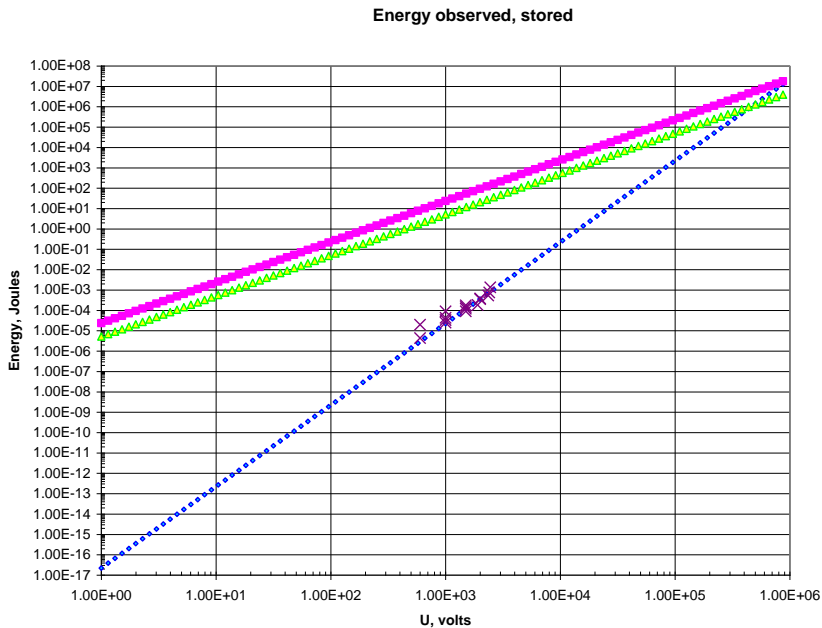

Fig. 2  Energy from capacitors, N$_2$ boil off, pendulum.

The energy transferred to the pendulum is mgh, while the stored energy is ½ C U², with C=46.86 µF.

The nitrogen thruster hypothesis requires that 22% of the energy stored in the capacitors goes into nitrogen gas energy. An alternate interpretation would be that the thruster is only 22% efficient in converting energy to thrust.

Choose a typical case for comparison, with capacitor voltage U=2000 volts, corresponding to a stored energy of ½ CU² of 93.3 Joules. The energy in the ϒ beam is related to momentum as E=pc, and the mass is defined as $m_y = E/c^2$.

Table 1. Energy of recoil from ceramic mount.

| Recoil | $N_2$ | ϒ |
|---|---|---|
| Momentum | 0.036 kg m/s | 0.036 kg m/s |
| Velocity | 575 m/s | $3*10^8$ m/s |
| Mass | 62.6 mg | 0.12 ng |
| Energy | 20.7 J | 10.8 MJ |

The interpretation of table 1 is similar to that of a thruster, which achieves optimum energy efficiency with comparable propellant and payload masses. Imparting 1 mJ of energy into the 640 gram pendulum requires 62 mg of nitrogen, and 20.4 Joules. The 0.12 nanograms of mass ascribed to the ϒ beam has much lower energy efficiency, requiring 10.8 MJ.

The curves in figure 2 are extrapolated to megavolts, to illustrate an argument by Poher about energy conservation. A hypothetical discharge of 200 kV would vaporize 700 grams of nitrogen, comparable to the mass of the ceramic mount. The almost linear relation between pendulum momentum and stored energy is valid only for a recoil mass much smaller than the ceramic mount mass.

According to conventional theories, the possibility that the ϒ beam accounts for the recoil momentum is ruled out on the basis of energy conservation. In principle the theory might be fixed up, by allowing the equivalent of the near field in electrodynamics. An analogy is the electric motor, in which virtual photons are invoked to explain forces among magnetic elements. The Poher argument about the role of vacuum energy is not based on convincing data.

## Tests with other materials

The signals reported were only observed with a two layer sample. The sintered material of layer S1 was a classical cuprate $YBa_2Cu_3O_{7-x}$, with a critical temperature of 90k. Cerium and samarium replaced 5% to 20% of the yttrium atoms for layer S2, lowering the critical temperature to 50k. A typical ceramic had a diameter of 17±0.5mm, a length of 23±0.5 mm and a mass of 21±0.5g.

Three other physical effects were observed during discharges into the ceramics: emission of sound, emission of light, and electrical signals from a triple capacitor and a double solenoid inside the Faraday cage.

Discharges have been recorded in other materials:
- Normal conductors made of aluminum, brass or copper, both at room temperature and in liquid nitrogen,
- Ceramics of the same type of material, but of a different chemical composition,
- In fully superconductive ceramics with no layers of different critical temperatures,

- In ceramics, where the two kinds of materials, used to fabricate the layers, were mixed together before the final sintering treatment,
- In piezoelectric and ferroelectric materials such as BaSrTiO3, PZT, and PLZT, at room temperature, with a 300±10 Ω resistance in parallel.

During discharges in these materials, none of the effects described previously were observed.

The physical effects disappeared when the ceramic layer S1 was not superconducting. Typically 30 minutes was required to achieve stable effects while cooling down by adding liquid nitrogen, or warming up after the LN2 had evaporated from the cryostat.

## Mechanical momentum transfer with thin film

Poher attributed the effects to conditions in the transition zone between layers S1 and S2. Thin film ceramics were constructed, to enhance the transition zone. A thin 30 μm layer of the S1 cuprate was spread over a copper foil and sintered. The copper support plays the role of the S2 layer. Two similar electrodes were joined by cryogenic glue or by insulated screws. Effects were found to be proportional to the film area, typically 25 cm$^2$.

Data obtained using the thin film ceramics were summarized as a relation between momentum transfer to the horizontal pendulum and capacitor voltage,

$$p = 2.4 \cdot 10^{-7} \frac{kg \cdot m}{s} \left( \frac{U}{1 volt} \right)^2 \qquad \text{eqn. 8.}$$

The coefficient in eqn. 8 is a factor of 22 larger than that found for the two layer ceramic. The relation is expressed in terms of energy transfer to the pendulum, using eqns. 4,5,6.

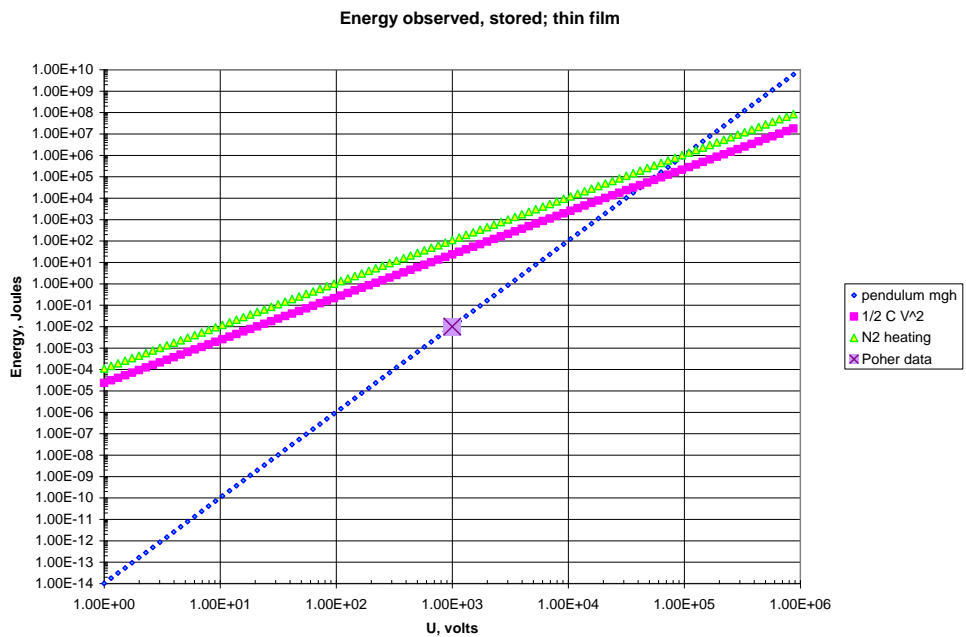

Fig. 3  Energy in capacitors, balance, nitrogen, using thin film ceramics.

Poher argues that the data imply that the apparatus comes close to violating energy conservation. However the analysis leading to this conclusion is traced to an error by Poher in applying eqns. 4,5,6 to convert momentum to energy. Figure 3 shows a crossover between pendulum energy and ½ C U$^2$ stored energy at 50 kilovolts. Poher notes problems with the stability of the cryostat for voltages above 1100 volts.

The nitrogen thruster model requires an efficiency of 400%, which is unphysical. Some other mechanism must be invoked, to explain typically 0.04% efficiency in transferring energy from the capacitor bank to the pendulum.

## Piezo detector measurements

Various detectors were placed in a Faraday cage below the target. Results from a piezoelectric detector placed 29 to 95 cm from the target are described.

A 0.687 gram mass was attached to the piezoelectric sensor to form an accelerometer. The detector was calibrated by the impact of a tiny body of known mass falling from several known heights. A typical set of runs produced accelerometer data in the form of sensor output in millivolts versus capacitor voltage. For a discharge voltage of 2900 volts, Poher interprets the piezo output of about 1.8 mV as an impulse of $8.8*10^{-8}$ kg m/s. Assuming a quadratic relation between impulse and capacitor voltage gives the relation

$$p_{pz} = 8.8 \cdot 10^{-8} \frac{kg \cdot m}{s} \left( \frac{U}{2900 volts} \right)^2 \qquad \text{eqn. 9.}$$

The Poher piezo detector data in the form of momentum versus voltage can be compared with similar data[5] from Podkletnov. The Podkletnov detector was a pendulum with an 18.5 gram mass. The Podkletnov high voltage pulse generator is based on 20 capacitors each 25 nf. The capacitors are charged to up to 100 kV in parallel, then switched to series. The source has a capacitance of C = 25/20 = 1.25 nf, and a maximum voltage V=2 MV. The energy in the capacitor bank is

$$E_{cap} = \tfrac{1}{2} CV^2 = 2.5 \text{ KJ} \qquad \text{eqn. 10.}$$

The current from the Podkletnov capacitor bank was measured as I=10$^4$ amps, implying a pulse width

$$\tau_{Pod} = CV/I = 250 \text{ nsec} \qquad \text{eqn. 11.}$$

For a potential of 2 MV, the momentum is p=0.007 kg m/s. Assuming that the momentum is carried by a massless field obeying E=pc, the energy required for p=0.007 kg m/s is

E = pc = 2.1 MJ            eqn. 12.

A theory requiring E=pc is not tenable.

Data from the two experiments are compared in terms of an acceleration field. The measured momentum is divided by the product of the detector mass and capacitor pulse width.

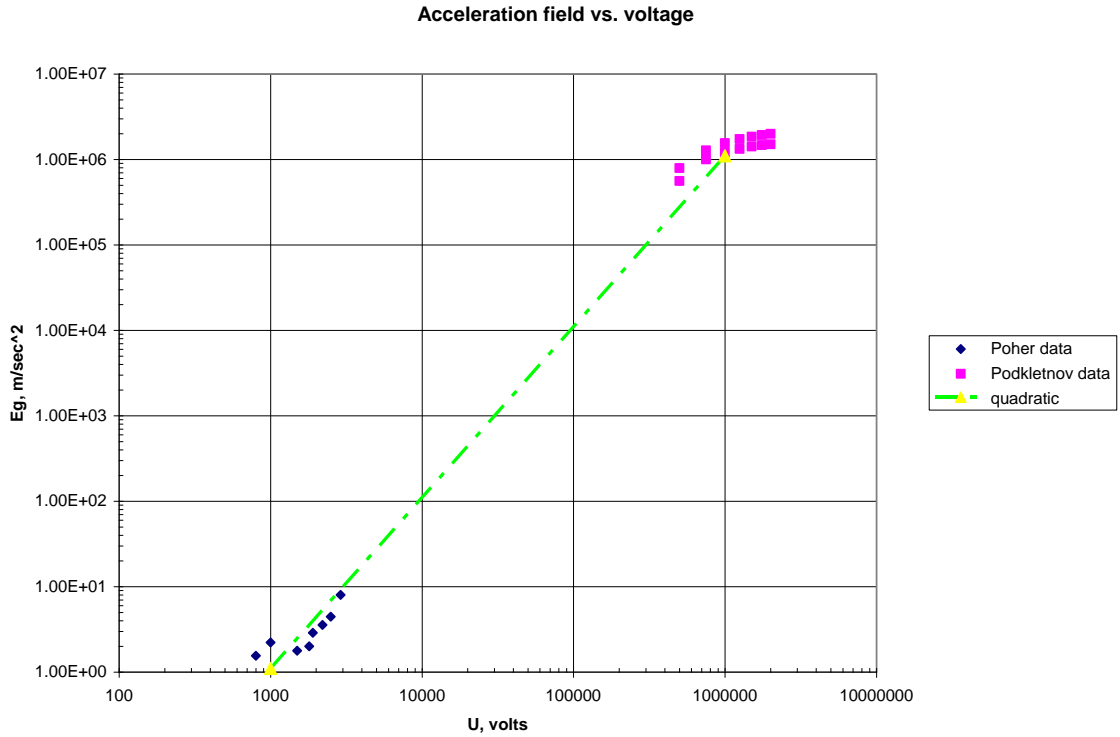

Fig. 4 Acceleration field sensed by remote detectors.
Parameters from the Poher and Podkletnov measurements are tabulated.

Table 2. Parameters of Poher, Podkletnov experiments

|  | Poher | Podkletnov |
|---|---|---|
| Accelerometer mass | 0.687 g | 18.5 g |
| Pulse width | 16 μsec | 250 nsec |
| Typical momentum | $4 \times 10^{-8}$ kg m/sec | 0.007 kg m/s |
| Acceleration | 3.6 m/sec$^2$ | $1.5 \times 10^6$ m/sec$^2$ |
| Beam energy, E=pc | 12 J | 2.1 MJ |

The data indicate order of magnitude agreement between Podkletnov data and an extrapolation of Poher results, assuming a quadratic relation between acceleration field and capacitor voltage.

## Spatial distribution of disturbances

Both Poher and Podkletnov report that the disturbance resembles a beam, with amplitude which is insensitive to distance from the ceramic, and a lateral extent comparable to the size of the ceramic disc. The detector response scales linearly with mass of the detector.

Poher describes data taken with a piezoelectric accelerometer, at a distance of 490 mm from the ceramic. The pendulum detectors in the Podkletnov measurements were 6 m to 150 m from the YBaCuO. Podkletnov reports that the force field is repulsive.

Podkletnov states[6] that the disturbance propagated in a well-collimated beam, with clean borders, having the same width as the superconducting emitter. The beam size was measured up to ca. 5 mm by means of special boards that receive an imprint under the action of small pressure, similar to those commercially available (for instance from Sensor Products, East Hanover).

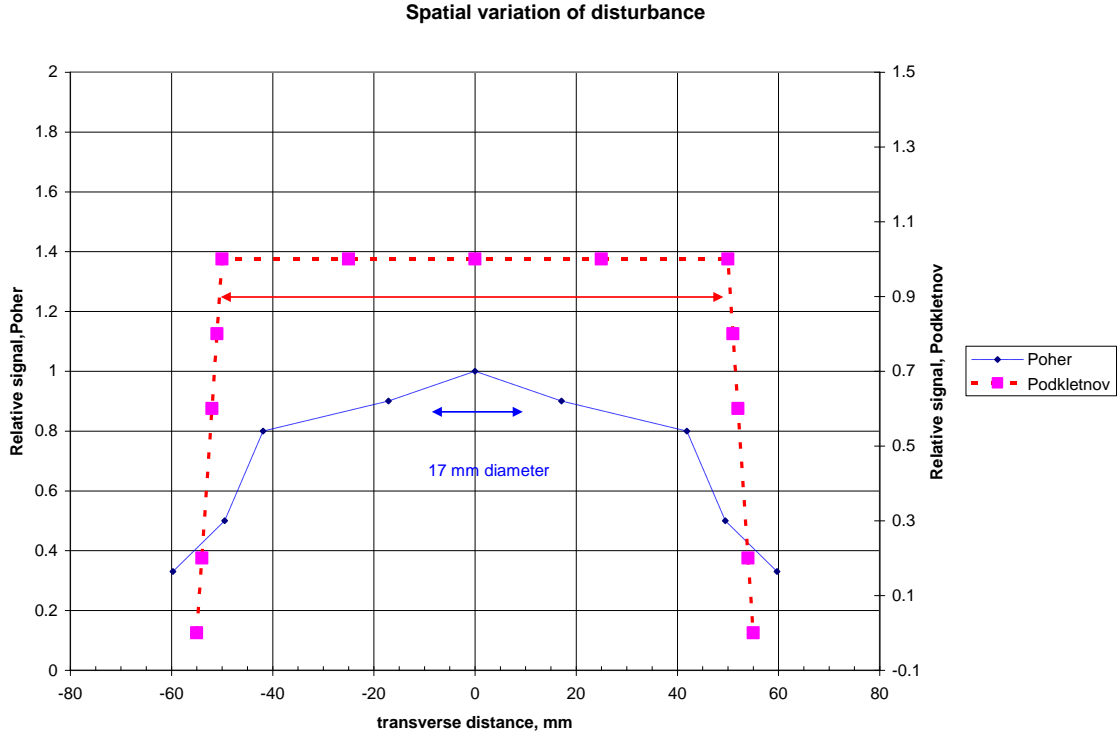

Fig. 5  Transverse profile of beam amplitude.

Superconducting emitters used by Podkletnov have diameters of 80 to 120 mm.

Podkletnov describes another experiment, in which the membrane of a microphone was tilted relative to the direction of propagation of the beam.  The data are consistent with longitudinal polarization.

## Complaints about experiment descriptions

Poher describes the calibration of the piezo detector in terms of output voltage versus force.  The conversion to impulse involves a timing parameter of about 1 msec, presumably a property of the accelerometer.  This detail is not explained.

Podkletnov describes the pulse width as $10^{-5}$ to $10^{-4}$ seconds, which disagrees with the 250 nsec in eqn. 11.

Both Poher and Podkletnov state that the momentum transfer is proportional to the detector mass.  However no data with masses other than 0.687 g and 18.5 g are shown.

## Phenomenological Dirac string description

Features of the data resemble the flux tube description[7] of meson spectroscopy.  The hypothesis borrowed from the flux tube description is that current through the YBaCuO boundary layer creates gravitoelectric Dirac monopoles, with a field described by a vector potential

$$\vec{C}(\vec{x}) = G_{enh} M \frac{1+\cos\theta}{r\sin\theta} \hat{\varphi} \qquad \text{eqn. 13,}$$

where M is the monopole mass, spherical coordinates (r,θ,$\varphi$) aligned with the symmetry axis of the YBaCuO. The coupling constant $G_{enh}$ is adjusted empirically to match data.

The acceleration field is the curl of the vector potential, which is singular along the +z axis. The Stokes theorem relates the average acceleration field to parameters of the model. The singular field occupies a tubular region surrounding the +z axis, with a radius equal to the YBaCuO radius. The average acceleration field in the flux tube is then

$$\langle \vec{E}_g \rangle = 4 G_{enh} M \frac{dN}{dA} \hat{z} \qquad \text{eqn. 14,}$$

where dN/dA is the number of Dirac strings per unit area. The detectors are insensitive to the thickness of the strings, assumed to be small compared with detector radii. The data are described with the following parameters:

| Parameter | Value | Description |
| --- | --- | --- |
| M | $9.1*10^{-31}$ kg | Electron mass |
| dN/dA | $1.4*10^{25}$ strings/m$^2$ | Density of Cooper pairs in a YBaCuO layer |
| $G_{enh}$ | $G_N * 3.2*10^{14} * (U/1000 \text{ volts})^2$ | Enhancement of Newtonian gravitational constant |

The Dirac string model describes the following features of the data:
- The acceleration field is independent of distance along the z axis;
- The transverse extent of the field is comparable to the YBaCuO radius;
- The force on a detector is proportional to the mass;
- The force is not appreciably affected by intervening materials.
- The force is repulsive, consistent with the polarity reversal inside and outside the Dirac string.

The electron mass is assigned to the monopole to avoid an expenditure of Mc$^2$ in creating the monopoles, which would otherwise lead to problems with energy conservation. Explaining the enhancement to the gravitational constant is beyond the scope of this review. Such an enhancement makes an appearance in theories[8] involving higher dimensions. Hints of monopole components of quarks and leptons have been claimed[9] in an analysis of high energy neutrino interactions. However the NuTeV anomaly can also be explained[10] by adjusting a few parameters of the Standard Model.

## Summary and comments

Poher hypothesizes that all of the effects observed with cold YBaCuO are explainable in terms of the Universon model. The Universon model invokes a force field directed in the direction of electron acceleration. Problems with energy conservation are attributed to vacuum energy. Poher admits that the source of the energy is unknown.

The momentum transfer to the Poher horizontal pendulum is much too large be consistent with propagation of a massless Universon beam with an E=pc energy momentum relation. An explanation involving vaporization of liquid nitrogen is plausible. Unlike metals, the YBaCuO materials are porous, allowing liquid nitrogen to be trapped. The LN2 hypothesis is that the mechanism resembles a thruster, with nitrogen gas momentum directed downward. The model explains the linear relation between momentum transfer and capacitor energy. An efficiency of 22% is consistent with data taken with a two layer ceramic.

A thin film ceramic gave much larger effects, requiring an unphysical 400% efficiency for the nitrogen thruster model. A more plausible model involving motion of liquid nitrogen rather than evaporation is beyond the scope of this review.

The Poher remote accelerometers pick up typically $10^{-8}$ kg-m/sec of momentum. Ascribing the momentum to a massless beam with the energy momentum relation E=pc leads to the result that 15% of the stored energy is required to produce the observed detector momenta. The accelerometers have a diameter of 6.25 mm, so that they intercept a small fraction (14%) of the 17 cm diameter transverse profile of the beam. Explaining the momentum transfer requires an implausible 100% efficiency for converting energy from the capacitor to the beam.

Podkletnov reports detector momenta 5 orders of magnitude larger than those reported by Poher. Attempting to account for the Podkletnov results encounters violations of energy conservation, assuming a beam with E=pc.

An extrapolation of Poher remote sensor data gives order of magnitude agreement with Podkletnov data.

If the experiments are correct, a theory must be constructed to allow an extended vacuum disturbance, describable as a flux tube. A phenomenological model involving Dirac string flux tubes summarizes features of the data.